\documentclass[12pt]{iopart}

\usepackage{graphicx}
\usepackage{bm}
\usepackage{color}

\begin{document}

\title[Dephasing in an Aharonov-Bohm interferometer]{Dephasing in an Aharonov-Bohm interferometer containing a lateral double quantum dot induced by coupling with a quantum dot charge sensor}

\author{T. Kubo$^{1,2}$, Y. Tokura$^{1,2}$, and S. Tarucha$^{1,3}$}

\address{$^1$Quantum Spin Information Project, ICORP, JST, Atsugi-shi, Kanagawa 243-0198, Japan\\
$^2$NTT Basic Research Laboratories, NTT Corporation, Atsugi-shi, Kanagawa 243-0198, Japan\\
$^3$Department of Applied Physics, University of Tokyo, Bunkyo-ku, Tokyo 113-8656, Japan}
\ead{kubo@will.brl.ntt.co.jp}
\begin{abstract}
We theoretically investigated the dephasing in an Aharonov-Bohm interferometer containing a lateral double quantum dot induced by coupling with a quantum dot charge sensor. We employed the interpolative 2nd-order perturbation theory to include the charge sensing Coulomb interaction. It is shown that the visibility of the Aharonov-Bohm oscillation of the linear conductance decreases monotonically as the sensing Coulomb interaction increases. In particular, for a weak sensing interaction regime, the visibility decreases parabolically, and it behaves linearly for a strong sensing interaction regime.

\end{abstract}

\maketitle

\section{Introduction}
Particle-wave duality is one of the most important concepts in quantum mechanics and provides the most impressive illustration of Bohr's complementarity principle \cite{bohr1,bohr2}. The wave characteristic arises only when the different possible paths that a particle can take are indistinguishable, even in principle (for example, see \cite{feynman}). By introducing a which-path detector, it is well-known that coupling with a charge sensor induces dephasing in an interferometer \cite{which-path,which-way}.

Mesoscopic systems are often used to study the interplay between the interference and dephasing of electrons. Recent nano-fabrication and low-temperature measuring techniques using semiconductors have enabled us to observe the various coherent effects of electrons such as the Aharonov-Bohm (AB) \cite{ab,yacoby,holleitner,hatano}, Fano \cite{kobayashi}, and Kondo effects \cite{sasaki,wilfred,kubo2}. The AB effect has been proven to be a convenient way of observing the interference fringe in mesoscopic systems since it provides an experimentally straightforward way of controling the phase. In interference experiments with an AB ring containing a QD, periodic modulation of the tunneling current as a function of the magnetic flux threading through the ring has been experimentally demonstrated \cite{yacoby,ab-exp1,ab-exp2}. This reflects the fact that the quantum phase coherence is maintained during the tunneling process through a QD. Moreover, controllable dephasing via a which-path detector has been demonstrated in a system with a quantum dot (QD) embedded in an AB ring \cite{which-path}. As regards the dephasing in a QD induced by coupling with a quantum point contact (QPC), theoretical studies have successfully explained the experimental results \cite{qpc-t1,qpc-t2,qpc-t3}. However, in this paper, we discuss the dephasing in an AB interferometer containing a lateral double quantum dot (DQD) induced by a coupling with a QD charge sensor. We consider the DQD system and use the QD charge sensor for the following reasons. Recently, the AB oscillations of a tunneling current passing through a lateral double quantum dot (DQD) system were observed by Holleitner \textit{et al}. \cite{holleitner} and Hatano \textit{et al}. \cite{hatano}. In such systems, we can detect the fresh AB effects through a quantum mehcanical superposition state such as the tunnel-coupled symmetric and antisymmetric states in a DQD \cite{kang,kubo,tokura}. In lateral DQD systems, the notion of coherent indirect coupling between two QDs via a reservoir is important. Previous theoretical work on the dephasing in an AB interferometer induced by coupling with a QD charge sensor has only discussed maximum coherent indirect coupling. In this paper, however, we study the dephasing when the coherent indirect coupling between two QDs in an interferometer is finite rather than maximal. Actually various experimental conditions correspond to such conditions. Moreover, according to the theory of dephasing caused by an electron-electron Coulomb interaction \cite{imry,stern,imry-book}, the dephasing rate is related to the charge fluctuation. In QD systems, the shot noise shows the intriguing behavior because of the many-body correlation effect under a finite source-drain bias voltage \cite{fujii}. Thus, we can expect to observe unconventional dephasing induced by coupling with a QD charge sensor. However, the many-body correlation in a QD charge sensor is not discussed in this paper since we deal with spinless electrons to focus on coherent charge transport.

This paper is organized as follows. In Sec. \ref{sec2} we employ the standard tunneling Hamiltonian formalism to describe an AB interferometer containing a DQD that couples to a QD charge sensor, and provide the formulation needed to calculate the transport properties. In particular, we introduce the notion of coherent indirect coupling between two QDs via a reservoir \cite{kubo,tokura}. In Sec. \ref{sec3} we calculate the nonequilibrium Green's functions within the framework of the interpolative 2nd-order nonequilibrium perturbation theory \cite{interpolative}. In Sec. \ref{sec4} we review the transport properties through an AB interferometer containing a lateral DQD. Section \ref{sec5} is devoted to numerical results of dephasing resulting from coupling with a QD charge sensor. And Sec. \ref{sec6} provides some concluding remarks.

\section{Model and formulation\label{sec2}}

\begin{figure}[htbp]
  \begin{center}
    \includegraphics[scale=0.6]{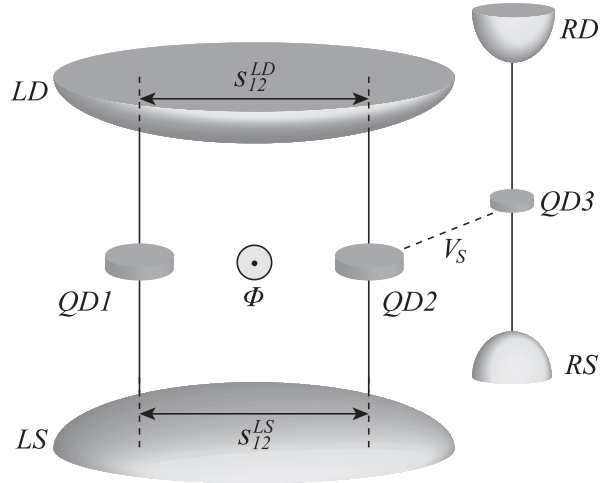}
  \end{center}
  \caption{Schematic diagram of an Aharonov-Bohm interferometer containing two quantum dots with a capacitively coupled quantum dot charge sensor. QD3 is capacitively coupled to QD2 and plays the role of charge sensor. We take account of the propagation of electrons in the reservoirs. $s_{12}^{\nu}$ is the propagation length, where $\nu\in LS,LD$. $V_S$ is a sensing interaction. $\Phi$ is the magnetic flux threading through an Aharonov-Bohm interferometer, which causes the Aharonov-Bohm effect.}
  \label{fig:system.eps}
\end{figure}
We consider an AB interferometer containing two QDs with a capacitively coupled QD charge sensor as shown in Fig. \ref{fig:system.eps}. In Fig. \ref{fig:system.eps}, two QDs ($QD1$ and $QD2$) couple to a common left source ($LS$) and left drain ($LD$) reservoirs, and the QD charge sensor ($QD3$) couples to the right source ($RS$) and right drain ($RD$) reservoirs. Moreover, we assume that the left and right reservoirs are completely separate. We neglect the spin degree of freedom, and only a single energy level in each QD is assumed to be relevant. We model this system with the Hamiltonian
\begin{eqnarray}
H=H_R+H_{DQD}+H_S+H_I+H_T,
\end{eqnarray}
where
\begin{eqnarray}
H_R=\sum_{\nu\in\{LS,LD,RS,RD\}}\sum_k\epsilon_{\nu k}{c_{\nu k}}^{\dagger}c_{\nu k}
\end{eqnarray}
describes the Fermi seas of noninteracting electrons in the $LS$, $LD$, $RS$, and $RD$ reservoirs. Here $\epsilon_{\nu k}$ is the electron energy with a wave number $k$ in a reservoir $\nu$, and the operator $c_{\nu k}$ (${c_{\nu k}}^{\dagger}$) annihilates (creates) an electron in the reservoir $\nu$. The Hamiltonian $H_{DQD}$ is
\begin{eqnarray}
H_{DQD}=\sum_{j=1}^2\epsilon_j{d_j}^{\dagger}d_j,
\end{eqnarray}
where $\epsilon_j$ is the single-particle energy level, and $d_j$ (${d_j}^{\dagger}$) annihilates (creates) an electron in the $j$-th QD ($j=1,2$). $H_S$ represents the QD charge sensor
\begin{eqnarray}
H_S=\epsilon_3{d_3}^{\dagger}d_3,
\end{eqnarray}
where $\epsilon_3$ is the energy level of QD3 and $d_3$ (${d_3}^{\dagger}$) is the annihilation (creation) operator of QD3. $H_I$ is the interaction between QD2 and QD3
\begin{eqnarray}
H_I=V_Sn_{22}n_{33}.
\end{eqnarray}
Here $V_S$ is the sensing interaction energy, namely the inter-dot Coulomb interaction energy between QD2 and QD3, and we introduce the following notation: $n_{jj}\equiv{d_j}^{\dagger}d_j$. $H_T$ is the tunneling Hamiltonian between the reservoirs and QDs
\begin{eqnarray}
H_T&=&\sum_k\left[t_{LSk}^{(1)}e^{i\frac{\phi}{4}}{c_{LSk}}^{\dagger}d_1+t_{LSk}^{(2)}e^{-i\frac{\phi}{4}}{c_{LSk}}^{\dagger}d_2\right.\nonumber\\
&&+t_{LDk}^{(1)}e^{-i\frac{\phi}{4}}{c_{LDk}}^{\dagger}d_1+t_{LDk}^{(2)}e^{i\frac{\phi}{4}}{c_{LDk}}^{\dagger}d_2\nonumber\\
&&\left. +t_{RSk}^{(3)}{c_{RSk}}^{\dagger}d_3+t_{RDk}^{(3)}{c_{RDk}}^{\dagger}d_3+\mbox{h.c.}\right]\nonumber\\
&\equiv&\sum_{\nu\in\{LS,LD\}}\sum_k\sum_{j=1}^2\left[t_{\nu k}^{(j)}(\phi){c_{\nu k}}^{\dagger}d_j+\mbox{h.c.} \right]\nonumber\\
&&+\sum_{\nu\in\{RS,RD\}}\sum_k\left[t_{\nu k}^{(3)}{c_{\nu k}}^{\dagger}d_3+\mbox{h.c.} \right]
\end{eqnarray}
where $t_{\nu k}^{(j)}$ are the tunneling amplitudes and real number. The factors $e^{\pm i\frac{\phi}{4}}$ indicate the effect of the magnetic flux ($\phi=2\pi\frac{\Phi}{\Phi_0}$ is an AB phase in the AB interferometer, where $\Phi$ is the magnetic flux threading through an AB interferometer as shown in Fig. \ref{fig:system.eps}, and $\Phi_0=\frac{h}{e}$ is the magnetic flux quantum).

The linewidth functions are defined by
\begin{eqnarray}
\Gamma_{ij}^{\nu}(\epsilon,\phi)=2\pi\sum_k{t_{\nu k}^{(i)}}^*(\phi)t_{\nu k}^{(j)}(\phi)\delta(\epsilon-\epsilon_{\nu k}).
\end{eqnarray}
Moreover, in the wide-band limit, we neglect the energy dependence of the linewidth functions. In our model, the left and right reservoirs are separate. As a result, $\Gamma_{13}^{\nu}=\Gamma_{23}^{\nu}=0$. Here we introduced the notation $M_{ij}$, which denotes the ($i,j$) matrix element of $3\times3$ matrix $\bm{M}$. The boldface notation indicates a $3\times3$ matrix whose basis is a localized state in each QD. Then the flux-dependent linewidth functions are given by
\begin{eqnarray}
\bm{\Gamma}^{S}(\phi)&=&\left(
  \begin{array}{ccc}
    \Gamma_{LS}   & \alpha_{LS}e^{-i\phi/2}\Gamma_{LS}   &  0  \\
    \alpha_{LS}e^{i\phi/2}\Gamma_{LS}   & \Gamma_{LS}   &  0  \\
    0   &  0  & \Gamma_{RS}   \\
  \end{array}
\right)\equiv\bm{\Gamma}^{LS}+\bm{\Gamma}^{RS},\\
\bm{\Gamma}^{D}(\phi)&=&\left(
  \begin{array}{ccc}
    \Gamma_{LD}   &  \alpha_{LD}e^{i\phi/2}\Gamma_{LD}  &  0  \\
    \alpha_{LD}e^{-i\phi/2}\Gamma_{LD}   &  \Gamma_{LD}  &  0  \\
    0   &  0  &  \Gamma_{RD}  \\
  \end{array}
\right)\equiv\bm{\Gamma}^{LD}+\bm{\Gamma}^{RD}
\end{eqnarray}
where
\begin{eqnarray}
\bm{\Gamma}^{LS}(\phi)&=&\left(
  \begin{array}{ccc}
    \Gamma_{LS}   & \alpha_{LS}e^{-i\phi/2}\Gamma_{LS}   & 0   \\
    \alpha_{LS}e^{i\phi/2}\Gamma_{LS}   &  \Gamma_{LS}  &  0  \\
    0   & 0   &  0  \\
  \end{array}
\right),\\
\bm{\Gamma}^{LD}(\phi)&=&\left(
  \begin{array}{ccc}
    \Gamma_{LD}   &  \alpha_{LD}e^{i\phi/2}\Gamma_{LD}  &  0  \\
    \alpha_{LD}e^{-i\phi/2}\Gamma_{LD}   &  \Gamma_{LD}  &  0  \\
    0   &  0  &  0  \\
  \end{array}
\right).
\end{eqnarray}
Here we introduced the coherent indirect coupling parameter $\alpha_{\nu}$, which characterizes the strength of the indirect coupling between QD1 and QD2 via the reservoir $\nu$ \cite{kubo}. In general, the coherent indirect coupling parameters $\alpha_{\nu}$ are a function of the propagation length $s_{12}^{\nu}$ as shown in Fig. \ref{fig:system.eps}. In DQD systems, in general, $|\alpha_{\nu}|\le 1$.

The linear conductances through a DQD and QD charge sensor are \cite{current}
\begin{eqnarray}
G_{DQD}&=&\frac{e^2}{h}\int_{-\infty}^{\infty}d\epsilon\left[-\frac{\partial f(\epsilon)}{\partial\epsilon} \right]\mbox{Tr}\left\{\bm{G}^r(\epsilon)\bm{\Gamma}^{LS}\bm{G}^a(\epsilon)\bm{\Gamma}^{LD} \right\},\\
G_S&=&\frac{e^2}{h}\int_{-\infty}^{\infty}d\epsilon\left[-\frac{\partial f(\epsilon)}{\partial\epsilon} \right]\mbox{Im}\left\{\frac{\Gamma_{RS}\Gamma_{RD}}{\Gamma_{RS}+\Gamma_{RD}}G_{33}^r(\epsilon) \right\},
\end{eqnarray}
where the retarded Green's function $\bm{G}^r(\epsilon)$ is the Fourier transform of
\begin{eqnarray}
G_{ij}^r(t,t')=-i\theta(t-t')\langle \{d_i(t),{d_j}^{\dagger}(t') \} \rangle,
\end{eqnarray}
and the advanced Green's function $\bm{G}^a(\epsilon)$ is obtained from the retarded Green's function: $\bm{G}^a(\epsilon)=[\bm{G}^r(\epsilon)]^{\dagger}$. Moreover the tunneling current through a QD charge sensor is
\begin{eqnarray}
I_S=-\frac{2e}{h}\int_{-\infty}^{\infty}d\epsilon[f_{RS}(\epsilon)-f_{RD}(\epsilon)]\mbox{Im}\left\{\frac{\Gamma_{RS}\Gamma_{RD}}{\Gamma_{RS}+\Gamma_{RD}}G_{33}^r(\epsilon) \right\},
\end{eqnarray}
and the population of the $j$th QD is given by
\begin{eqnarray}
\langle n_j\rangle=\int\frac{d\epsilon}{2\pi i\hbar}G_{jj}^{-+}(\epsilon).
\end{eqnarray}
Here the lesser Green's function $\bm{G}^{-+}(\epsilon)$ is the Fourier transform of
\begin{eqnarray}
G_{ij}^{-+}(t,t')=i\langle {d_j}^{\dagger}(t')d_i(t) \rangle,
\end{eqnarray}
and $f_{\nu}(\epsilon)$ is the Fermi-Dirac distribution function of the reservoir $\nu$ defined as
\begin{eqnarray}
f_{\nu}(\epsilon)=\frac{1}{1+e^{(\epsilon-\mu_{\nu})/k_BT}},
\end{eqnarray}
where $\mu_{\nu}$ is the electrochemical potential of the reservoir $\nu$, and $f(\epsilon)$ is the equilibrium Fermi-Dirac distribution function, namely $\mu_{\nu}=0$. In the following discussions, we consider the situation when the DQD is always contained in the linear response, and the source and drain reservoirs coupled with the QD charge sensor have electrochemical potentials $\mu_{RS}=\mu+eV_{SD}/2$ and $\mu_{RD}=\mu-eV_{SD}/2$ with the source-drain bias voltage $V_{SD}$, and $\mu=0$.

\section{Nonequilibrium perturbation theory\label{sec3}}
To calculate the observable quantities as discussed in the previous section, we need the nonequilibrium Green's functions. The formal Schwinger-Keldysh perturbation theory \cite{schwinger,keldysh,lifshitz} with respect to the Coulomb interaction $V_S$ provides a good description only at the particle-hole symmetric point because of the failure of the Friedel-Langreth sum rule \cite{friedel,langreth} and the violation of the current conservation \cite{2nd,interpolative}. Moreover, far from the particle-hole symmetric point, the 2nd-order self-energy does not yield the atomic limit $\Gamma/V_S\to 0$ \cite{interpolative}. However, we would like to discuss a wider regime such as the QD energy and sensing Coulomb interaction dependences of the linear conductance. Thus, we employ the interpolative 2nd-order nonequilibrium perturbation theory, which satisfies the current conservation condition and fulfills the Friedel-Langreth sum rule \cite{interpolative}.

In accordance with the interpolative 2nd-order nonequilibrium perturbation theory \cite{interpolative}, we consider the following effective one-electron Hamiltonian:
\begin{eqnarray}
H_{eff}&=&\sum_{\nu\in\{LS,LD,RS,RD\}}\left(\epsilon_{\nu k}+\mu_{\nu,eff} \right){c_{\nu k}}^{\dagger}c_{\nu k}+\sum_{j=1}^3\epsilon_{j,eff}{d_j}^{\dagger}d_j\nonumber\\
&&+\sum_{\nu\in\{LS,LD\}}\sum_k\sum_{j=1}^2\left[t_{\nu k}^{(j)}(\phi){c_{\nu k}}^{\dagger}d_j+\mbox{h.c.} \right]\nonumber\\
&&+\sum_{\nu\in\{RS,RD\}}\sum_k\left[t_{\nu k}^{(3)}{c_{\nu k}}^{\dagger}d_3+\mbox{h.c.} \right],\label{eff-ham}
\end{eqnarray}
where $\mu_{LS,eff}=\mu_{LD,eff}=0$, $\epsilon_{1,eff}=\epsilon_1$, and $\mu_{RS,eff}$, $\mu_{RD,eff}$, $\epsilon_{2,eff}$, and $\epsilon_{3,eff}$ are determined by imposing self-consistency in the QD populations and the tunneling current. For the effective Hamiltonian (\ref{eff-ham}), we can calculate the effective one-electron Green's functions
\begin{eqnarray}
\bm{g}_{eff}^{r}(\epsilon)&=&\left(
  \begin{array}{ccc}
    \frac{\epsilon-\epsilon_1}{\hbar}+\frac{i}{2}\Gamma_{11}   & \frac{i}{2}\Gamma_{12}   & 0   \\
    \frac{i}{2}\Gamma_{21}   &  \frac{\epsilon-\epsilon_{2,eff}}{\hbar}+\frac{i}{2}\Gamma_{22}  &  0  \\
    0   &  0  &  \frac{\epsilon-\epsilon_{3,eff}}{\hbar}+\frac{i}{2}\Gamma_{33}  \\
  \end{array}
\right)^{-1},\\
\bm{g}_{eff}^a(\epsilon)&=&[\bm{g}_{eff}^{r}(\epsilon)]^{\dagger},\\
\bm{g}_{eff}^{-+}(\epsilon)&=&i\sum_{\nu\in\{LS,LD,RS,RD\}}f_{\nu,eff}(\epsilon)\bm{g}_{eff}^r(\epsilon)\bm{\Gamma}^{\nu}\bm{g}_{eff}^a(\epsilon),\\
\bm{g}_{eff}^{+-}(\epsilon)&=&-i\sum_{\nu\in\{LS,LD,RS,RD\}}[1-f_{\nu,eff}(\epsilon)]\bm{g}_{eff}^r(\epsilon)\bm{\Gamma}^{\nu}\bm{g}_{eff}^a(\epsilon),
\end{eqnarray}
where $f_{\mu,eff}(\epsilon)$ is the effective Fermi-Dirac distribution function defined as
\begin{eqnarray}
f_{\mu,eff}(\epsilon)=\frac{1}{1+e^{(\epsilon-\mu_{\nu,eff})/k_BT}}.
\end{eqnarray}
Using these Green's functions, the 2nd-order retarded self-energies are
\begin{eqnarray}
\Sigma_{22}^{r(2)}(\epsilon)&=&\left(\frac{V_S}{2\pi} \right)^2\int\frac{dE_1}{\hbar}\int\frac{dE_2}{\hbar}\left[g_{22,eff}^r(E_1)g_{33,eff}^{+-}(E_2)g_{33,eff}^{-+}(E_1+E_2-\epsilon) \right.\nonumber\\
&&+g_{22,eff}^{-+}(E_1)g_{33,eff}^{r}(E_2)g_{33,eff}^{+-}(E_1+E_2-\epsilon)\nonumber\\
&&\left. +g_{22,eff}^{-+}(E_1)g_{33,eff}^{+-}(E_2)g_{33,eff}^{a}(E_1+E_2-\epsilon)\right],\\
\Sigma_{33}^{r(2)}(\epsilon)&=&\left(\frac{V_S}{2\pi} \right)^2\int\frac{dE_1}{\hbar}\int\frac{dE_2}{\hbar}\left[g_{33,eff}^r(E_1)g_{22,eff}^{+-}(E_2)g_{22,eff}^{-+}(E_1+E_2-\epsilon) \right.\nonumber\\
&&+g_{33,eff}^{-+}(E_1)g_{22,eff}^{r}(E_2)g_{22,eff}^{+-}(E_1+E_2-\epsilon)\nonumber\\
&&\left. +g_{33,eff}^{-+}(E_1)g_{22,eff}^{+-}(E_2)g_{22,eff}^{a}(E_1+E_2-\epsilon)\right].
\end{eqnarray}
To improve the solution so that the self-enegy yields an appropriate atomic limit for $\Gamma/V_S\to 0$, we introduce the following interpolative self-enegies \cite{interpolative}:
\begin{eqnarray}
\Sigma_{22}^r(\epsilon)&=&\frac{\Sigma_{22}^{r(2)}(\epsilon)}{1-\frac{(1-\langle n_3\rangle)V_S+\epsilon_2-\epsilon_{2,eff}}{\langle n_3\rangle(1-\langle n_3\rangle)(V_S)^2}\Sigma_{22}^{r(2)}(\epsilon)},\\
\Sigma_{33}^r(\epsilon)&=&\frac{\Sigma_{33}^{r(2)}(\epsilon)}{1-\frac{(1-\langle n_2\rangle)V_S+\epsilon_3-\epsilon_{3,eff}}{\langle n_2\rangle(1-\langle n_2\rangle)(V_S)^2}\Sigma_{33}^{r(2)}(\epsilon)}.
\end{eqnarray}
Then, the full retarded Green's function is given by
\begin{eqnarray}
\bm{G}^r(\epsilon)=\left[\left(\bm{g}^r(\epsilon) \right)^{-1}-\bm{\Sigma}^r(\epsilon) \right]^{-1},
\end{eqnarray}
where
\begin{eqnarray}
\bm{g}^r(\epsilon)&=&\left(
  \begin{array}{ccc}
     \frac{\epsilon-\epsilon_1}{\hbar}+\frac{i}{2}\Gamma_{11}  & \frac{i}{2}\Gamma_{12}   & 0   \\
     \frac{i}{2}\Gamma_{21}  & \frac{\epsilon-\epsilon_2-V_S\langle n_3\rangle}{\hbar}+\frac{i}{2}\Gamma_{22}   &  0  \\
     0  &  0  & \frac{\epsilon-\epsilon_3-V_S\langle n_2\rangle}{\hbar}+\frac{i}{2}\Gamma_{33}   \\
  \end{array}
\right)^{-1},\\
\bm{\Sigma}^r(\epsilon)&=&\left(
  \begin{array}{ccc}
    0   &  0  & 0   \\
    0   &  \Sigma_{22}^r(\epsilon)  & 0   \\
    0   &  0  &  \Sigma_{33}^r(\epsilon)  \\
  \end{array}
\right),
\end{eqnarray}
and $\epsilon_2+V_S\langle n_3\rangle$ and $\epsilon_3+V_S\langle n_2\rangle$ are the Hartree dressed QD energies. Similarly, we can calculate the lesser Green's function $\bm{G}^{-+}(\epsilon)$. Finally, we impose self-consistency in the QD populations and the tunneling current. From the self-consistent equations below, we can determine the unknown parameters $\mu_{RS,eff}$, $\mu_{RD,eff}$, $\epsilon_{2,eff}$, and $\epsilon_{3,eff}$:
\begin{eqnarray}
\langle n_j\rangle&=&\int\frac{d\epsilon}{2\pi i\hbar}g_{jj,eff}^{-+}(\epsilon)=\int\frac{d\epsilon}{2\pi i\hbar}G_{jj}^{-+}(\epsilon),\label{self-pop}\\
I_{\nu}&=&\frac{ie}{2\pi}\int\frac{d\epsilon}{\hbar}\mbox{Tr}\left\{\bm{\Gamma}^{\nu}\left[\bm{g}_{eff}^{-+}(\epsilon)+f_{\nu,eff}(\epsilon)\left(\bm{g}_{eff}^r(\epsilon)-\bm{g}_{eff}^a(\epsilon) \right) \right] \right\}\nonumber\\
&=&\frac{ie}{2\pi}\int\frac{d\epsilon}{\hbar}\mbox{Tr}\left\{\bm{\Gamma}^{\nu}\left[\bm{G}^{-+}(\epsilon)+f_{\nu}(\epsilon)\left(\bm{G}^r(\epsilon)-\bm{G}^a(\epsilon) \right) \right] \right\},
\end{eqnarray}
where $I_{\nu}$ is the current from the reservoir $\nu$ ($\nu\in RS,RD$) to the QD charge sensor, and Eq. (\ref{self-pop}) has to be employed for $j=2,3$ since there is finite sensing Coulomb interaction only between QD2 and QD3. When we focus on the linear response for the QD charge sensor, the self-consistency of the current is not required.

\section{Transport properties of a DQD without coupling with a QD charge sensor\label{sec4}}
\begin{figure}[htbp]
  \begin{center}
    \includegraphics[scale=0.5]{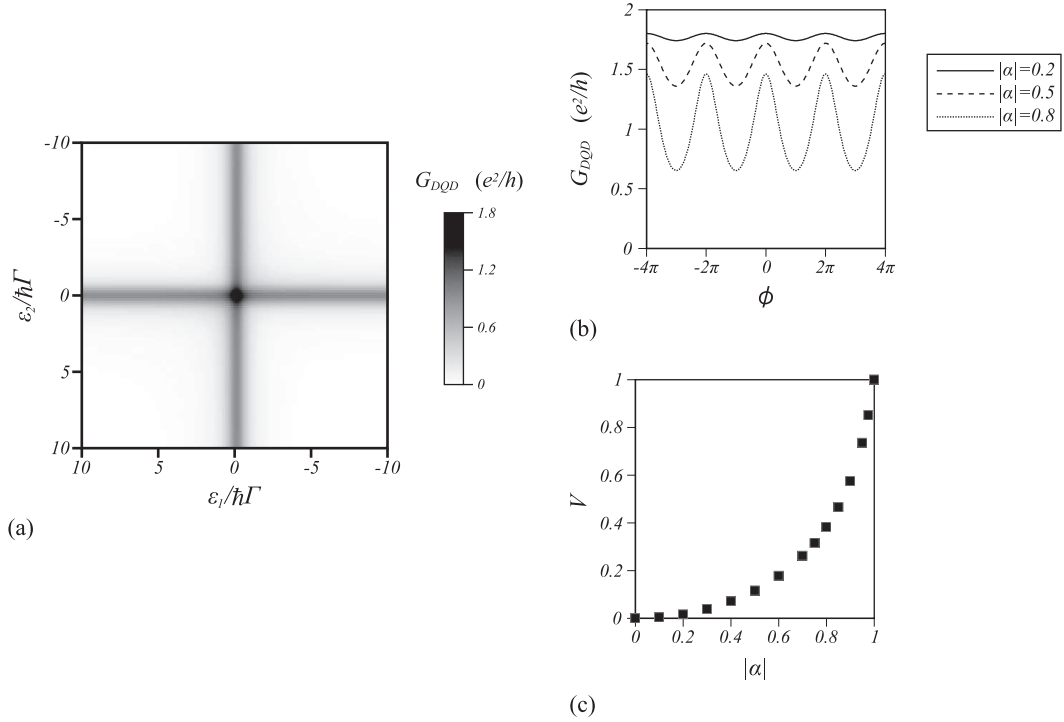}
  \end{center}
  \caption{Linear conductance through a DQD. (a) $G_{DQD}$ as a function of $\epsilon_1$ and $\epsilon_2$ when $\phi=0$, $\alpha=0.5$, and $k_BT/\hbar\Gamma=0.1$. (b) AB oscillations of $G_{DQD}$ for various $\alpha$ when $\epsilon_1=\epsilon_2=0$ (at the crossing point in (a)). (c) Coherent indirect coupling parameter dependence of the visibility of the AB oscillations in $G_{DQD}$.}
  \label{fig:noninteracting-AB.eps}
\end{figure}

Before we discuss the dephasing in an AB interferometer, here we briefly review the transport properties of an AB interferometer containing a DQD without coupling with a QD charge sensor, namely $V_S=0$. In the following discussions, for clarity, we consider a symmetric condition where $\alpha_{LS}=\alpha_{LD}\equiv\alpha$ and $\Gamma_{\nu}=\Gamma/2$. The linear conductance through a DQD is shown in Fig. \ref{fig:noninteracting-AB.eps}(a) as a function of $\epsilon_1$ and $\epsilon_2$ (charge stability diagram) when $\phi=0$, $\alpha=0.5$, and $k_BT/\hbar\Gamma=0.1$. We observe two conductance peaks at $\epsilon_1=0$ and $\epsilon_2=0$ caused by the resonant tunneling through each QD. In Fig. \ref{fig:noninteracting-AB.eps}(b), we plot the magnetic flux dependences of the linear conductance through a DQD, namely AB oscillations, for various $\alpha$ values at fixed QD energies of $\epsilon_1=\epsilon_2=0$ (crossing point of two conductance peaks). The solid, broken, and dotted lines indicate $|\alpha|=0.2$, $|\alpha|=0.5$, and $|\alpha|=0.8$, respectively. In all cases, the AB oscillation period is $2\pi$. We find that the visibility of the AB oscillations increases monotonically as $|\alpha|$ increases, where the visibility is defined by
\begin{eqnarray}
V=\frac{G_{max}-G_{min}}{G_{max}+G_{min}}.
\end{eqnarray}
Here $G_{max}$ and $G_{min}$ correspond to the maximum value at $\phi=2n\pi$ and the minimum value at $\phi=(2n+1)\pi$, where $n$ is an integer. In Fig. \ref{fig:noninteracting-AB.eps}(c), we show the coherent indirect coupling parameter dependence of the visibility of the AB oscillations in the linear conductance through a DQD.

\section{Dephasing induced by coupling with QD charge sensor\label{sec5}}
In this section, we discuss the dephasing in an AB interferometer containing a DQD induced by coupling with a QD charge sensor. In our calculation, we find that the real part of the retarded self-energy provides the renormalization of the QD energy level from the expression of the retarded Green's function. In contrast, the imaginary part of the retarded self-energy provides the relaxation except for the escape rate to the reservoirs, which is described by the bare linewidth function $\Gamma$. We are interested in the effects of the latter. Therefore, in our calculations, we always compensate for the effects of the level shift, and we focus on AB oscillations at the crossing point between two conductance peaks. In this section, we concentrate on the situation where $\alpha=0.5$ and $k_BT/\hbar\Gamma=0.1$.

\subsection{Dephasing in linear regime}
\begin{figure}[htbp]
  \begin{center}
    \includegraphics[scale=0.6]{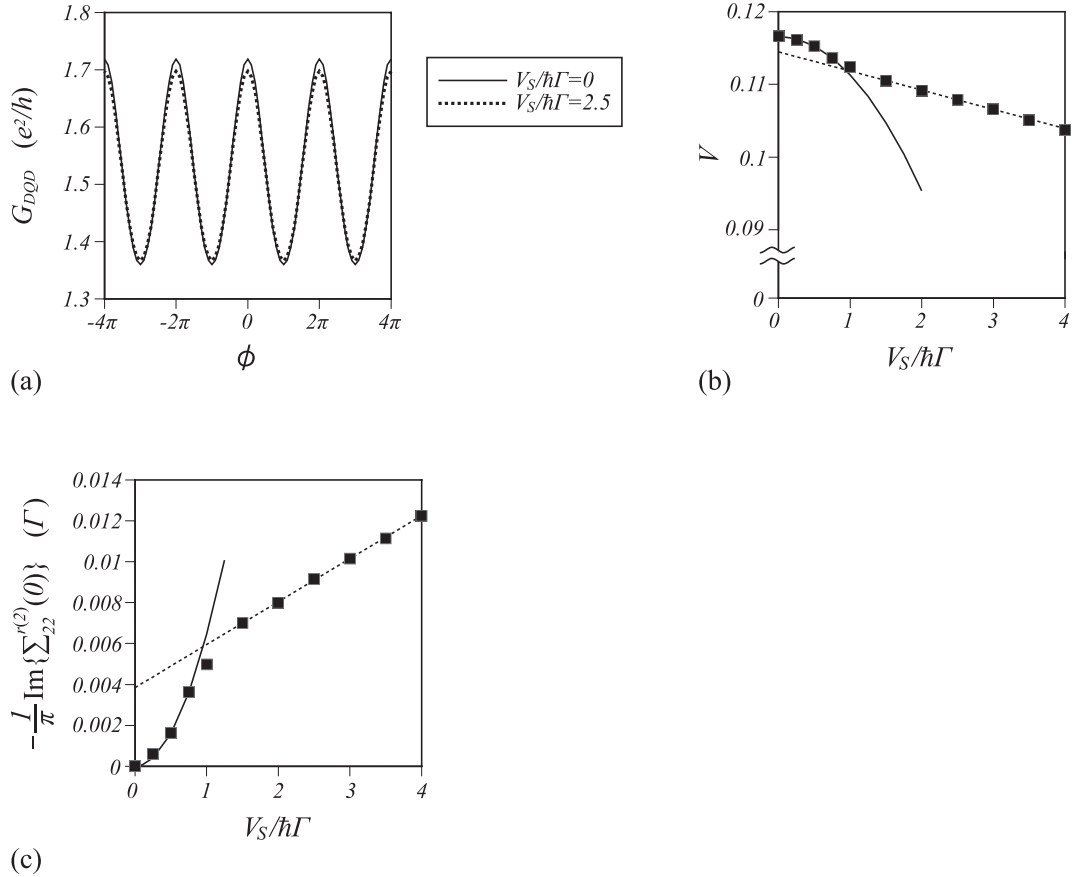}
  \end{center}
  \caption{AB oscillations of the conductance and its visibility in a linear regime for $\epsilon_3=0$. (a) AB oscillations of the linear conductance for $V_S=0$ and $V_S/\hbar\Gamma=2.5$. (b) Sensing Coulomb interaction dependence of the visibility of the AB oscillations in the linear conductance. The solid and dotted lines correspond to parabolic and linear fitting lines. (c) Relaxation rate as a function of the sensing Coulomb interaction. The solid and dotted lines correspond to parabolic and linear fitting lines.}
  \label{fig:linear.eps}
\end{figure}

Here we consider linear transport. In Fig. \ref{fig:linear.eps}, for $\epsilon_3=0$, we plot the AB oscillations of the linear conductance through a DQD and  the sensing Coulomb interaction dependence of their visibility $V$.

As shown in Fig. \ref{fig:linear.eps}(a), the visibility $V$ decreases as the sensing Coulomb interaction becomes stronger. According to Fig. \ref{fig:linear.eps}(b), the visibility decreases monotonically as the sensing Coulomb interaction increases. However, the reduction behavior changes when the strength of the sensing Coulomb interaction is $V_S\sim\hbar\Gamma$. For a weak interaction regime ($V_S<\hbar\Gamma$), the visibility decreases parabolically (fitted by the solid line in Fig. \ref{fig:linear.eps}(b)). In a strong interaction regime ($V_S>\hbar\Gamma$), the visibility exhibits linear dependence (fitted by the dotted line in Fig. \ref{fig:linear.eps}(b)). In Fig. \ref{fig:linear.eps}(c), we plot the relaxation rate $-\mbox{Im}\left\{\Sigma_{22}^{r(2)}(0)\right\}$ as a function of the sensing interaction. In the weak interaction regime, the relaxation rate increases parabolically as the sensing interaction increases, and exhibits a linear increase in a strong interaction regime. Thus it seems that the relaxation rate is related in some way to the visibility. However, why the visibility and the relaxation rate exhibit linear dependences remains an open problem.

\begin{figure}[htbp]
  \begin{center}
    \includegraphics[scale=0.6]{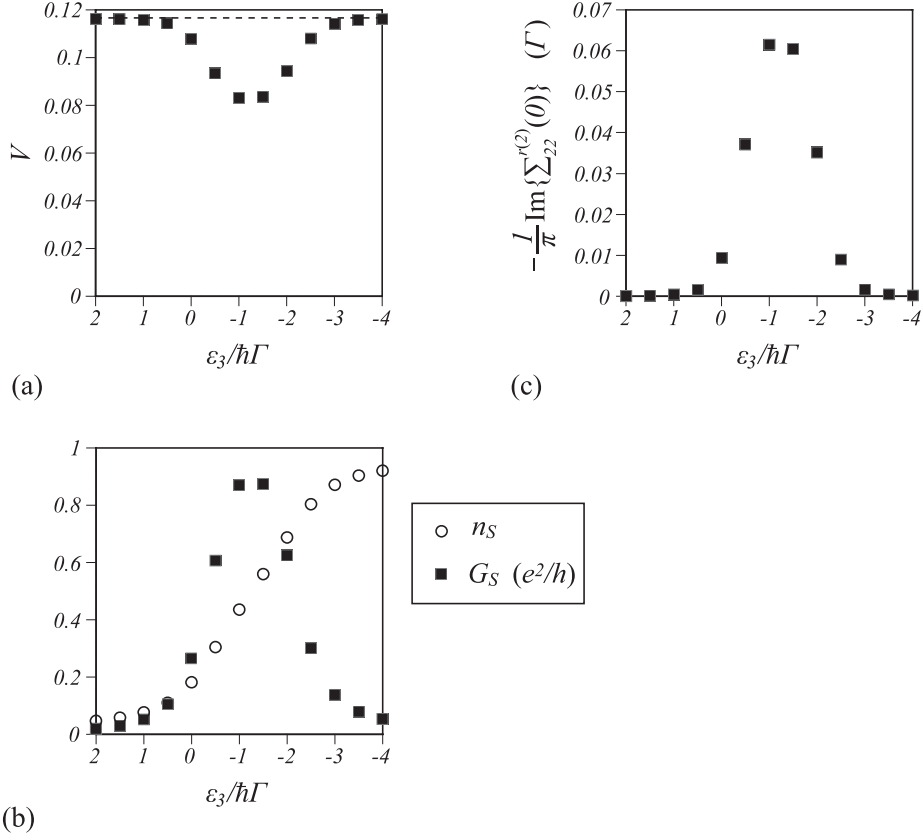}
  \end{center}
  \caption{Energy dependence of the QD charge sensor on the visibility of the AB oscillations in the linear conductance through a DQD for $V_S/\hbar\Gamma=2.5$. (a) Visibility of the AB oscillations. The dotted line corresponds to the visibility without a sensing Coulomb interaction. (b) Linear conductance through a QD charge sensor and its population. Squares and circles indicate the linear conductance and the population of the QD charge sensor, respectively. (c) Relaxation rate due to the coupling with a QD charge sensor.}
  \label{fig:linear-eps3.eps}
\end{figure}

Next we study the energy dependence of the QD charge sensor on the visibility of the AB oscillations in the linear conductance through a DQD. As shown in Fig. \ref{fig:linear-eps3.eps}(a), we plot the energy dependence of the QD charge sensor on the visibility when $\alpha=0.5$ and $V_S/\hbar\Gamma=2.5$. We find that the visibility has a single dip structure. Away from the minimum point, the visibility gradually approaches the visibility value without a sensing Coulomb interaction. To understand this behavior, we investigate the transport property and the population of the QD charge sensor as shown in Fig. \ref{fig:linear-eps3.eps}(b). According to Fig. \ref{fig:linear-eps3.eps}, the position of the conductance peak through a QD charge sensor is shifted from $\epsilon_3=0$ as a result of the Coulomb interaction $V_S$ between QD2 and QD3. This shift is dominantly due to the Hartree contribution $V_S\langle n_2\rangle /\hbar\Gamma\sim1.25$. We find that the visibility is minimal at the position of the conductance peak through the QD charge sensor. Moreover, we show the relaxation rate $-\mbox{Im}\left\{\Sigma_{22}^{r(2)}(0)\right\}$ caused by the sensing Coulomb interaction in Fig. \ref{fig:linear-eps3.eps}(c). When the relaxation rate is at its maximum value, the visibility becomes minimum. Therefore, similar to the dephasing induced by coupling with the QPC charge sensor, the visibility is related to the relaxation rate in our problem.

\subsection{Dephasing in nonlinear regime}
\begin{figure}[htbp]
  \begin{center}
    \includegraphics[scale=0.6]{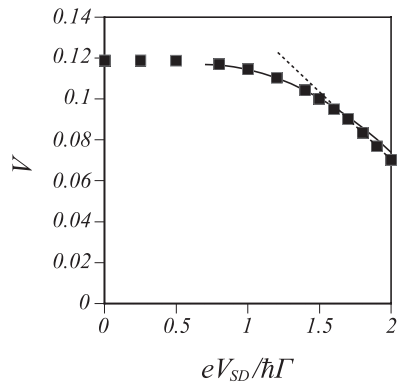}
  \end{center}
  \caption{Source-drain bias voltage dependence of the visibility of the AB oscillations of the tunneling current through a DQD for $V_S/\hbar\Gamma=2.5$. The solid and dotted lines indicate the parabolic and linear fitting lines, respectively.}
  \label{fig:nonlinear.eps}
\end{figure}

Here we discuss nonlinear transport, namely the case when the QD charge sensor is under a finite source-drain bias voltage. When $\epsilon_3=0$ and $V_S/\hbar\Gamma=2.5$, the visibility is approximately constant for a low source-drain bias voltage regime ($eV_{SD}\ll \hbar\Gamma$) as shown in Fig. \ref{fig:nonlinear.eps}. In an intermediate regime ($eV_{SD}\sim\hbar\Gamma$), the visibility exhibits the parabolic dependence, and finally it behaves linearly for a high source-drain bias voltage regime ($eV_{SD}\gg \hbar\Gamma$). In a nonlinear regime, it appears that our result is qualitatively similar to the result obtained for a QPC charge sensor \cite{which-path}. However, in the QPC charge sensor, the visibility is almost constant when the QPC source-drain bias voltage is lower than the temperature, and the visibility behaves linearly when the QPC source-drain bias voltage is higher than the temperature. Our calculation shows that the visibility behavior changes in $eV_{SD}\sim\hbar\Gamma$ in the QD charge sensor. This suggests that the QPC and QD charge sensors may have essentially different dephasing mechanisms.

\section{Conclusions\label{sec6}}
Here we summarize the main results of this study. We examined the dephasing in an AB interferometer containing a lateral double quantum dot induced by coupling with a QD charge sensor using the interpolative 2nd-order nonequilibrium perturbation theory. As regards the linear rensponse, we found that the visibility decreases parabolically for a weak interaction regime and behaves linearly in a strong interaction regime. Moreover, we found that the visibility may relate to the relaxation rate from two kinds of calculations. In a nonlinear regime, the visibility is almost constant for a low source-drain bias voltage and behaves linearly in the intermediate region, and finally decreases linearly at a high source-drain bias voltage. In this paper, we assumed that the sensing interaction is relatively strong. The calculation in a weak interaction regime, for example $V_S<\hbar\Gamma$, and the relationship to the shot noise through a QD charge sensor will be reported elsewhere \cite{preparation}. We have already fabricated lateral triple QD systems using vertical QDs \cite{amaha}. We can realize such systems as discussed in this paper by using one of three QDs as the charge sensor.

\section*{Acknowledgments}
We thank A. Aharony, O. Entin-Wohlman, M. Eto, S. Sasaki, and T. Hatano for valuable discussions and useful comments.


\section*{References}
\begin{thebibliography}{10}
\bibitem{bohr1} Bohr N, 1928 Nature \textbf{121} 580
\bibitem{bohr2} Heisenberg W, 1930 \textit{The Physical Principles of the Quantum Theory} (Chicago Press)
\bibitem{feynman} Feynman R P, Leighton R B, and Sands M, 1965 \textit{The Feynman Lectures on Physics Volume 3} (Addison-Wesley Pub.)
\bibitem{which-path} Buks E, Schuster R, Heiblum M, Mahalu D, and Umansky V, 1998 Nature \textbf{391} 871
\bibitem{which-way} D\"{u}rr S, Nonn T, and Rempe G, 1998 Nature \textbf{395} 33
\bibitem{ab} Aharonov Y and Bohm D, 1959 Phys. Rev. \textbf{115} 485
\bibitem{yacoby} Yacoby A, Heiblum M, Mahalu D, and Shtrikman H, 1995 Phys. Rev. Lett. \textbf{74} 4047
\bibitem{holleitner} Holleitner A W, Decker C R, Qin H, Eberl K, and Blick R H, 2001 Phys. Rev. Lett. \textbf{87} 256802
\bibitem{hatano} Hatano T, Stopa M, Izumida W, Yamaguchi T, Ota T, and Tarucha S, 2004 Physica E (Amsterdam) \textbf{22} 534
\bibitem{kobayashi} Kobayashi K, Aikawa H, Katsumoto S, and Iye Y, 2002 Phys. Rev. Lett. \textbf{88} 256806
\bibitem{sasaki} Sasaki S, 2000 Nature \textbf{405} 764
\bibitem{wilfred} W G van der Wiel, 2000 Science \textbf{289} 2105
\bibitem{kubo2} Kubo T, Tokura Y, and Tarucha S, 2008 Phys. Rev.B \textbf{77} 041305(R)
\bibitem{ab-exp1} Schuster W, Buks E, Heiblum M, Mahalu D, Umansky V, and Shtrikman H, 1997 Nature (London) \textbf{385} 417
\bibitem{ab-exp2} Ji Y, Heiblum M, Sprinzak D, Mahalu D, and Shtrikman H, 2000 Science \textbf{290} 779
\bibitem{qpc-t1} Aleiner I L, Wingreen N S, and Meir Y, 1997 Phys. Rev. Lett. \textbf{79} 3740
\bibitem{qpc-t2} Levinson Y, 1997 Europhys. Lett. \textbf{39} 299
\bibitem{qpc-t3} Hackenbroich G, 2001 Phys. Rep. \textbf{343} 463
\bibitem{kang} Kang K and Cho S Y, 2004 J. Phys. Condens. Matter \textbf{16} 117
\bibitem{kubo} Kubo T, Tokura Y, Hatano T, and Tarucha S, 2006 Phys. Rev. B \textbf{74} 205310
\bibitem{tokura} Tokura Y, Nakano H, and Kubo T, 2007 New J. Phys. \textbf{9} 113
\bibitem{imry} Stern A, Aharonov Y, and Imry Y, 1990 Phys. Rev. A \textbf{41} 3436
\bibitem{stern} Stern A, Aharonov Y, and Imry Y, 1991 \textit{Quantum Coherence in Mesoscopic Systems, NATO ASI Series No 254} (Plenum Press, New York)
\bibitem{imry-book} Imry Y, 1997 \textit{Introduction to Mesoscopic Physics} (Oxford, New York: Oxford University Press)
\bibitem{fujii} Fujii T, 2007 J. Phys. Soc. Jpn. \textbf{76} 44709
\bibitem{interpolative} Levy Yeyati A, Mart\'{i}n-Rodero A and Flores F, 1993 Phys. Rev. Lett. \textbf{71} 2991
\bibitem{current} Meir Y, Wingreen N S, 1992 Phys. Rev. Lett. \textbf{68} 2512
\bibitem{schwinger} Schwinger J, 1961 J. Math. Phys. \textbf{2} 407
\bibitem{keldysh} Keldysh L V, 1964 Zh. Eksp. Teor. Fiz. \textbf{47} 1515 [1965 Sov. Phys. JETP \textbf{20} 1018]
\bibitem{lifshitz} Lifshitz E M and Pitaevskii L P, \textit{Physical Kinetics} (Landau and Lifshitz Course of theoretical physics, v.X.) (Oxford, New York: Pergamon Press)
\bibitem{friedel} Friedel J, 1958 Nuovo Cimento Suppl. \textbf{7} 287
\bibitem{langreth} Langreth D C, 1966 Phys. Rev. \textbf{150} 516
\bibitem{2nd} Hershfield S, Davies J H and Wilkins J W, 1992 Phys. Rev. B \textbf{46} 7046
\bibitem{preparation} Kubo T, Tokura Y, and Tarucha S, in preparation
\bibitem{amaha} Amaha S, Hatano T, Tamura H, Kubo T, Teraoka S, Tokura Y, Austing D G, and Tarucha S, 2010 Physica E \textbf{42} 899
\end{thebibliography}
\end{document}